\documentclass[prd,twocolumn,floats,superscriptaddress,eqsecnum,
floatfix]{revtex4}

\usepackage{graphics, graphicx}
\usepackage{epsfig}
\usepackage{amsmath,amssymb}
\usepackage{color}

\pdfoutput=1

\newcommand{\be}{\begin{equation}}
\newcommand{\ee}{\end{equation}}
\newcommand{\ba}{\begin{eqnarray}}
\newcommand{\ea}{\end{eqnarray}}

\newcommand{\lap}{\bigtriangleup}
\newcommand{\eq}[1]{Eq.(\ref{#1})}

\newcommand{\hh}{\, ,\hspace{0.6cm}}
\newcommand{\hhh}{\, ,\hspace{0.4cm}}

\newcommand{\ins}[1]{{\mbox{\tiny #1}}}

\begin{document}

\title{Scalar and electromagnetic fields of static sources in higher dimensional
Majumdar-Papapetrou spacetimes}

\author{Valeri Frolov}%
\email[e-mail: ]{vfrolov@ualberta.ca}
\affiliation{Theoretical Physics Institute, Department of
Physics,
University of Alberta,\\
Edmonton, Alberta, Canada T6G 2E1  and \\
Yukawa Institute for Theoretical Physics, Kyoto University, Kyoto, Japan
}
\author{Andrei Zelnikov}%
\email[e-mail: ]{zelnikov@ualberta.ca}
\affiliation{Theoretical Physics Institute, Department of Physics,
University of Alberta,\\
Edmonton, Alberta, Canada T6G 2E1
}

\begin{abstract}
We study massless scalar and electromagnetic fields from static sources in a
static higher dimensional spacetime. Exact expressions for static Green's
functions for such problems are obtained in the background of the
Majumdar-Papapetrou solutions of the Einstein-Maxwell equations.
Using this result we calculate the force between two scalar or electric charges
in the presence of one or several extremally charged black holes in equilibrium
in the higher dimensional spacetime.
\end{abstract}


\maketitle

\section{Introduction}

There exists a variety of interesting physical problems that require study of a
test electromagnetic field in a curved spacetime. For such problems one uses a
test field
approximation: The field propagates in a fixed gravitational background and its
backreaction on
the metric is neglected. This approximation is often used for the study
of electric
and magnetic fields created by sources near a black hole. Black hole
electrodynamics
is an important part of the membrane paradigm \cite{MEMPAR} and
has important applications in black hole astrophysics.

In the present paper we consider the static field of a pointlike
charge placed in the black hole vicinity. In a local frame such a field close to
the charge is radial and similar to the field of a charge in a flat space.
However the action of gravity modifies the field at far distance from the
charge. This modification can be easily `explained' as the gravitational
attraction of the energy-stress distribution associated with the test
field\footnote{It is also well known that Maxwell equations in a curved
spacetime are equivalent to Maxwell in the media. In particular, a static
gravitational field plays the role of media with electric permittivity and
magnetic permeability $\epsilon=\mu=1/\sqrt{|g_{tt}|}$ \cite{LL}}.

The simplest case is a situation when an electric charge is at rest in a uniform
gravitational field. The corresponding electric field can be easily obtained by
writing the corresponding Li\'{e}nard-Wiechert  potential for a uniformly
accelerated charge in the Rindler coordinates.
In 1921 Fermi \cite{fermi} used this approach to study how the homogeneous
gravitational field affects the self-energy of an electric charge.
A similar problem of the self energy for charged particles at rest in the
vicinity of a neutral and charged black holes was discussed more recently in
\cite{SmithWill:1980,FrolovZelnikov:1982,Lohiya:1982,ChoTsokarosWisseman:2007}.
The remarkable fact is
that the solutions for scalar massless and electric field created by a pointlike
charge in the Schwarzschild and Reissner-Nordstr\"{o}m metrics can be obtained
in
an explicit analytical form \cite{Copson:1928,LL,Linet:1976,Linet:1977}. This
result is a consequence of the generic property of these four dimensional
metrics: They can be written in the form of the Weyl metric. If the axis of the
symmetry is chosen so that it passes through the position of the charge, the
corresponding equations for the static field for such a source are effectively
reduced to the equations in a flat 3D space. Gibbons and Warnick
\cite{GibbonsWarnick:2009} demonstrated that the exact
static scalar and electromagnetic Green functions in the four-dimensional
Schwarzschild and Reissner-Nordstr\"om  spacetimes can be also obtained by
reducing the problem to calculations in optical metrics. Unfortunately such
methods do
not work for higher dimensional generalizations of the spherically symmetric
black holes. However recently Linet \cite{Linet:05} obtained explicit
exact solutions for
the field of a pointlike particle in the vicinity of higher dimensional
extremally
charged Reissner-Nordstr\"om black hole.

In this paper we obtain exact solution
of this problem in a wide class of physically interesting metrics
and provide a far going generalization of Linet's results.

The aim of this paper is to demonstrate that in the higher dimensional case
there exist wide class of physically interesting metrics where
the massless scalar and electric field equations allow exact solutions for
pointlike charges. This class includes so called Majumdar-Papapetrou solutions
of the Einstein-Maxwell equations.  The corresponding solutions  describe one or
several extremally charged higher dimensional black holes in equilibrium.
They are supersymmetric and saturate the Bogomol'nyi bound.
These solutions are widely discussed in the string theory (see e.g.
\cite{Ortin,Maeda}).
The $D-$dimensional Majumdar-Papapetrou solution is static and its spatial  part
is
conformal to the $(D-1)-$dimensional Euclidean metric. We demonstrate that this
property allows one to reduce static scalar and electric problems to similar
problems in a flat space.
\footnote{Let us notice that general relativistic charged fluids in
Majumdar-Papapetrou metrics and their generalizations were studied in the paper
\cite{LemosZanchin:2009}.}

The paper is organized as follows. In Section 2 we recall some properties of the
Majumdar-Papapetrou solutions and discuss two special cases: a single higher
dimensional extremal Reissner-Nordstr\"{o}m black hole, and such a black hole in
a space with compact extra dimension.
This material is well known. We collect here the results, basically in order to
fix notations we use in the main text. In Sections 3 and 4 we obtain explicit
expressions for scalar and electric field of a point charge in the
Majumdar-Papapetrou geometry. We also analyze interesting special cases and
discuss how the presence of extremally charged black holes modifies the
interaction force between charges in these spacetimes.


\section{Majumdar-Papapetrou solutions of Einstein-Maxwell equations}

\subsection{General form of the Majumdar-Papapetrou metric}

The higher dimensional Einstein-Maxwell action is\footnote{We put the speed of
light $c=1$ and
later on also the D-dimensional gravitational constant $G^{(D)}=1$.}
\begin{equation}\begin{split}
S&=S_\ins{g}+S_\ins{em}\,,
\\
S_\ins{g}&={1\over 16\pi G^{(D)}}\int d^{D}x \sqrt{-g}\,R,
\\
S_\ins{em}&=-{1\over 16\pi }\int d^{D}x
\sqrt{-g}\,F_{\mu\nu}F^{\mu\nu}+\int d^{D}x
\sqrt{-g}\,A_\mu J^\mu \,.
\end{split}\end{equation}

In $D=n+3$ dimensions the solution which describes the metric of a set of
extremally charged black holes at rest can be written in the form
\be\label{MP}
ds^2=-U^{-2}\,dt^2+U^{2/n}\delta_{ab}\,dx^a dx^b\, .
\ee
The corresponding static electric field is given by the vector potential
\be\label{calA}
{\cal A}_{\mu}=\sqrt{{n+1\over 2n}}\,U^{-1}\,\delta^0_{\mu}\,.
\ee
This is an exact solution of the Einstein-Maxwell
equations if the function $U$ satisfies the equation
\be\label{lapU}
\lap U=0\, ,
\ee
that is, it is a harmonic function. We denote by Greek indices the spacetime
coordinates and use Latin indices $a,b,\dots$ or bold face fonts for spatial
coordinates $x^\mu\equiv(t,x^a)\equiv(t,\boldsymbol{x})$. The
$(n+2)-$dimensional Laplacian in \eq{lapU} is defined in accordance to the flat
spatial metric
\be
\lap\equiv \delta^{ab}\partial_a\partial_b\,.
\ee
We also use here the flat metric to define the norm of the spatial vector
\be\label{abs}
|\boldsymbol{x}|\equiv \sqrt{\delta_{ab}\,x^a x^b}\,.
\ee
The `coordinate distance' between spatial points
$\boldsymbol{x}$ and $\boldsymbol{x}'$ then reads
$|\boldsymbol{x}-\boldsymbol{x}'|$ .

For a special choice of the solution
\be\label{U}
U=1+\sum_k {M_k\over \rho_k^n}\hh \rho_k=|\boldsymbol{x}-\boldsymbol{x}_k|\,,
\ee
the metric \eq{MP} describes multiple black holes in equilibrium, when
the gravitational attraction between them is exactly compensated by
the electromagnetic repulsion. These metrics are the higher dimensional 
generalization \cite{Myers:1986} of the
Majumdar-Papapetrou metrics.
The metric with $k=1,\ldots ,N$ describes $N$ extremally charged black holes in
a
static equilibrium. Here $\boldsymbol{x}_k$ is the spatial position of the
$k$-th extremal black hole.
The function $U$ obeys the homogeneous equation \eq{lapU} everywhere outside
these
points.
When these points are included one has
\be
\lap U =-{4\pi^{1+{n\over 2}}\over \Gamma\left({n\over 2}\right)}\sum_k
M_k\,\delta^{n+2}(\boldsymbol{x}-\boldsymbol{x}_k)\, .
\ee
These $\delta-$functions are localized on the horizons. They correspond to the
effective charge distributions 
\be\begin{split}
\sqrt{-g}\,{\cal J}^0&=-\sqrt{n+1\over 2n}{\lap U\over 4\pi}\\
&=\sqrt{n+1\over
2n}{\pi^{n\over 2}\over \Gamma\left({n\over 2}\right)}\sum_k
M_k\,\delta^{n+2}(\boldsymbol{x}-\boldsymbol{x}_k)\, .
\end{split}\ee
on the horizons of the charged black holes of the Majumdar-Papapetrou spacetime.

\subsection{Special cases}

\subsubsection{Higher dimensional extremal Reissner-Nordstr\"om black hole}

A simplest case of the Majumdar-Papapetrou metric is an
 an extremal Reissner-Nordstr\"om black hole
\be\begin{split}\label{RN}
 ds^2&=-U^{-2}\,dt^2+U^{2/n}\,dr^2+r^2\,d\Omega_{n+1}^2\,,\\
 U&=\left(1-{r_{\ins{g}}^n\over r^n}\right)^{-1}\,.
\end{split}\ee
Here $r_{\ins{g}}$ is the gravitational radius of the black hole and
$d\Omega_{n+1}^2$ is the metric on a unit \mbox{$(n+1)$-dimensional} sphere
\be
d\Omega_{n+1}^2=d\theta_{n+1}^2+\sin^2\theta_{n+1}d\Omega_{n}^2\, .
\ee
We shall use notations
\be
\theta=\theta_{n+1}\hh \phi=\theta_1\, .
\ee
The angles $\theta_{j>1}$ change in the interval $(0,\pi)$, while
$\phi$ changes in the interval $(0,2\pi)$.

The vector potential ${\cal A}_\mu$ of the electric field is
\be
{\cal A}_{\mu}=\sqrt{{n+1\over 2n}}\,\left(1-{r_{\ins{g}}^n\over r^n}\right)
\,\delta^0_{\mu}\,.
\ee
At infinity the potential does not vanish but asymptotically approaches a
pure
gauge solution.

After the coordinate transformation
\be\label{rho}
\rho^n=r^n-r_{\ins{g}}^n
\ee
the Reissner-Nordstr\"om metric takes the form, where the spatial part is
conformally flat
\be\begin{split}\label{rnrho}
 ds^2&=-U^{-2}\,dt^2+U^{2/n}\,\left(d\rho^2+\rho^2\,d\Omega_{n+1}^2\right)\,,\\
 U&=1+{r_{\ins{g}}^n\over\rho^{n}}\, .
\end{split}\ee
In this particular case of a single black hole these coordinated are called
isotropic. It is convenient to introduce the `coordinate distance' between two
points in
the metric \eq{rnrho} which is just the distance defined by the flat spatial
geometry $R(\boldsymbol{x},\boldsymbol{x'})=|\boldsymbol{x}-\boldsymbol{x'}|$
(see \eq{abs}).

\subsubsection{Compactified extremally charged black hole}

The  Majumdar-Papapetrou metric \eq{MP} can be used for description 
of a spacetime of a single black hole but in a compactified spacetime 
with the spatial topology of a cylinder \cite{Myers:1986}.
The idea is simple.
Consider at first a black hole in the space with the topology of a cylinder
and which is periodic in one direction, e.g., $L^a$ with the coordinate period
$L=|\boldsymbol{L}|$. This spacetime is equivalent to the multi black
hole metric where all black holes are aligned in one direction with an equal
distance between them and have the same masses. This is evidently a particular
case of the generic higher dimensional Majumdar-Papapetrou metric with the
function
\be
U(\boldsymbol{x})=1+M\sum_{k=-\infty}^{\infty} {1\over \rho_k^n}\hh
\rho_k=|\boldsymbol{x}-\boldsymbol{x}_\ins{BH}-k\boldsymbol{L}|\, .
\ee
Without loss of generality one can always put the black hole in the coordinate
origin $\boldsymbol{x}_\ins{BH}=0$, and choose $L^a=(0,\dots,0,L)$. Then
\be\begin{split}
\rho_k&=\sqrt{[(z-kL)^2+\ell^2]} \hh
 z\equiv x^{n+2} \,,\\
 \ell^2&\equiv\delta_{ij}x^ix^j \hh
i,j=(1,\dots,n+1)\, .
\end{split}\ee 
For example in five dimensions $(n=2)$ the summation leads to
\be
U(\boldsymbol{x})=1+M\,{\pi\over \ell L}\,{\sinh{(2\pi \ell /
L)}\over\cosh{(2\pi
\ell /L)}-\cos{(2\pi z/ L)}}\,.
\ee
In any odd dimensions one can easily derive a more complicated but similar
expression in terms of elementary functions. Namely, if $j=n/2$ is integer then
\be
U(\boldsymbol{x})=1-M\,{(-1)^j\over j!}\,\left({1\over 2\ell}{\partial \over
\partial\ell}\right)^j\left[\ln\left(\cosh{2\pi
\ell \over L}-\cos{2\pi z\over L}\right)\right]\,.
\ee
In even dimensions (odd $n$) there is no simple expression for this sum.

\section{Static massless scalar field in Majumdar-Papapetrou spacetimes}

\subsection{Massless scalar field with sources}

The best way to deal with this problem is to
start with the total action for the particles carrying a scalar charge. It
consists of the scalar
field action, the action of the massive particle itself, and the interaction
term
\begin{equation}\begin{split}
S&=S_\ins{sc}+S_\ins{m}+S_\ins{int}\,,
\\
S_\ins{sc}&=-{1\over 8\pi}\int d^{D}x \sqrt{-g}\,
\varphi^{;\mu}\varphi_{;\mu}\,,
\\
S_\ins{m}&=-\int d\tau\,m\sqrt{-u^{\mu}u_{\mu}}\,,
\\
S_\ins{int}&=\int d^{D}x\,\sqrt{-g}\,\varphi(x) J(x)\,.
\end{split}\end{equation}

Consider two pointlike scalar charges $q$ and $q'$
located at points $\boldsymbol{y}$ and $\boldsymbol{y'}$, respectively, in
the spacetime with the metric \eq{MP}.
The  source describing a pointlike scalar charge moving along the worldline
$y^{\mu}(\tau)$ is
\begin{equation}\begin{split}
J(x)=q\,\int d\tau\,\sqrt{-u^{\mu}u_{\mu}}\,\delta^D(x^{\mu},{y}^{\mu})\,,
\end{split}\end{equation}
where $\tau$ is its proper time and
$\delta^D(x^{\mu},{y}^{\mu})=(-g)^{-1/2}\delta^D(x^{\mu}-{y}^{\mu})$ is the
covariant D-dimensional $\delta$-function $(D=n+3)$. In  this
expression the factor $\sqrt{-u^{\mu}u_{\mu}}$ is equal to 1 on the equations of
motion, but for an arbitrary off-shell trajectory it is to be considered a
functional of the path. This factor is necessary for the consistency of the
variation procedure. In order to calculate the force exerted by one scalar
charge to another, one has to keep in mind peculiar properties of the scalar
field.

Variation of this action over the particle path gives the equation of motion.
For a particle of mass $m$ and a scalar
charge $q$ we obtain
\be
{d\over d\tau}p_{\mu}=f_{\mu}.
\ee
Here 
\be
p_{\mu}=(m-q\phi)\,u_{\mu}
\ee
is the canonical momentum of the scalar particle.
The tricky point
is that the effective inertial mass of the scalar charge depends on the scalar
field  \cite{ChiuHoffmann:1964}
\be
m_\ins{eff}=m-q\varphi(y)
\ee
and, hence, is not constant in spacetime because in a general case the
scalar field $\varphi$ is not homogeneous. The force acting on the scalar charge
is 
\be
f_{\mu}=q\,\varphi_{;\mu }\,.
\ee
If the scalar field $\varphi$ is created by another static pointlike charge $q'$
located at the spatial \mbox{point $\boldsymbol{y}'$}
\begin{equation}\begin{split}
J'(\boldsymbol{x})&=q'\,{\sqrt{-g_{00}(\boldsymbol{x})}\over\sqrt{-g(\boldsymbol
{x})}}\,
\delta^{n+2}(\boldsymbol{x}-{ \boldsymbol{y}'})\\
&=q\,U^{-1-{2\over
n}}(\boldsymbol{x})\,\delta^{n+2}(\boldsymbol{x}-{\boldsymbol{y}'}).
\end{split}\end{equation}
then it can be expressed in terms of the Green function
\be\label{varphi}
\varphi(\boldsymbol{y})=4\pi q'\,\sqrt{-g_{00}(\boldsymbol{y}')}\,{\cal
G}(\boldsymbol{y},{\boldsymbol{y}'})\,.
\ee

\bigskip

\subsection{Field of a scalar point charge}

Let us consider a pointlike scalar charge in a $(n+3)-$dimensional static
spacetime with the metric
\be\label{mm}
ds^2=-U^2 dt^2+H^2 dl^2\, .
\ee
Here $dl^2$ is a flat $(n+2)-$dimensional metric
\be
dl^2=\delta_{ab}dx^a dx^b\hh a,b=1,\ldots, n+2\, ,
\ee
and $U$ and $H$ are functions of spatial coordinates $x^a$.
The massless scalar field equation is of the form
\be
\Box \varphi=-4\pi J\hh
\Box\equiv g^{\mu\nu}\nabla_\mu\nabla_\nu\, .
\ee
We focus our attention on a static field $\varphi$ generated by a static source
$J$. It obeys the equation
\be\label{eqq}
\Delta^{(n+2)} \varphi=-4\pi J\,,
\ee
where
\be
\Delta^{(n+2)}={U\over H^{n+2}} \delta^{ab}\partial_a\left(
U^{-1}H^n\partial_b\right)\, .
\ee
For a special case of Majumdar-Papapetrou spacetime, when $H=U^{1/n}$ the
operator $\Delta^{(n+2)}$ is
proportional for the $(n+2)-$dimensional flat Laplace operator
$\lap=\delta^{ab}\partial_a\partial_b$ and the equation
\eq{eqq} takes the form
\be
\lap\varphi=-4\pi U^{2/n}J\, .
\ee
The metric \eq{mm} in this case takes the form \eq{MP}.

The static Green function for $(n+2)-$dimensional Laplace operator is a solution
of the equation
\be
\lap{\cal
G}(\boldsymbol{x},\boldsymbol{x'})=-\delta^{n+2}(\boldsymbol{x}-\boldsymbol{x'}
)\, .
\ee
Using the following two relations
\be
\lap \left[{1\over \rho^n}\right]=-{4\pi^{1+{n\over 2}}\over \Gamma\left({n\over
2}\right)}\,\delta^{n+2}(\boldsymbol{x})
=-n{1\over \rho^{n+1}}\,\delta(\rho)\,,
\ee
\be
\delta^{n+2}(\boldsymbol{x})={\Gamma\left(1+{n\over 2}\right)\over
2\pi^{1+{n\over 2}}}\,{1\over \rho^{n+1}}\,\delta(\rho)\,,
\ee
one can easily obtain the Green function ${\cal
G}(\boldsymbol{x},\boldsymbol{x'})$
\be\label{Green}
{\cal G}(\boldsymbol{x},\boldsymbol{x'})={\Gamma\left({n\over 2}\right)\over
4\pi^{1+{n\over
2}}}\, \cdot {1\over R^n}\,,
\ee
where
\be
R(\boldsymbol{x},\boldsymbol{x'})=|\boldsymbol{x}-\boldsymbol{x'}|\, .
\ee
The function $R(\boldsymbol{x},\boldsymbol{x'})$ is the coordinate distance
\eq{abs} between points $\boldsymbol{x}$ and
$\boldsymbol{x'}$.

Therefore the solution for the scalar field corresponding to the generic static
scalar source $J$ reads
\be\begin{split}\label{phix}
\varphi(\boldsymbol{x})&=4\pi\int d^{n+2}x\, \sqrt{-g(\boldsymbol{x}')} \,
J(\boldsymbol{x}')\,{\cal G}(\boldsymbol{x},\boldsymbol{x'})\\
&=4\pi\int d^{n+2}x\,U^{2/n}(\boldsymbol{x}') \, J(\boldsymbol{x}')\,{\cal
G}(\boldsymbol{x},\boldsymbol{x'})
\end{split}\ee

\bigskip

\subsection{Force between two pointlike scalar charges}

Thus the force takes the form $f_{\mu}=(0,f_a)$ where
\be
f_{a}=-{n\,\Gamma\left({n\over 2}\right)\over \pi^{n\over 2}}
\,{1\over U(\boldsymbol{y}')}\,{qq'\over
R^{n+2}(\boldsymbol{y},\boldsymbol{y}')}
(y^b-y'{}^b)\,\delta_{ab} .
\ee
One can see that the force is attractive and $f^a\sim (y^a-y'{}^a) $,  i.e.,
when written in the metric \eq{MP}, it is directed exactly to the position of
the
charge $q'$. The invariant absolute value of the force
$|f|=\sqrt{f_\mu f^\mu}$ is
\be
|f|={n\,\Gamma\left({n\over 2}\right)\over \pi^{n\over 2}}
\,{1\over U^{1\over n}(\boldsymbol{y}) U(\boldsymbol{y}')}\,{qq'\over
R^{n+1}(\boldsymbol{y},\boldsymbol{y}')}.
\ee

\bigskip

\subsection{Special cases}

\subsubsection{Scalar field near higher dimensional extremally charged black
hole}

For a single black hole the higher dimensional
Majumdar-Papapetrou metric \eq{MP} reduces to the higher dimensional
version of the extremal Reissner-Nordstr\"om metric \eq{rnrho} in
isotropic coordinates. In curvature coordinates it is given by
\eq{RN} with the function
\be
U=1+{r^n_\ins{g}\over\rho^n}=\left( 1-{r^n_\ins{g}\over r^n}  \right)^{-1}\,.
\ee

\begin{figure}[h]
\includegraphics[width=6cm, height=4.5cm]{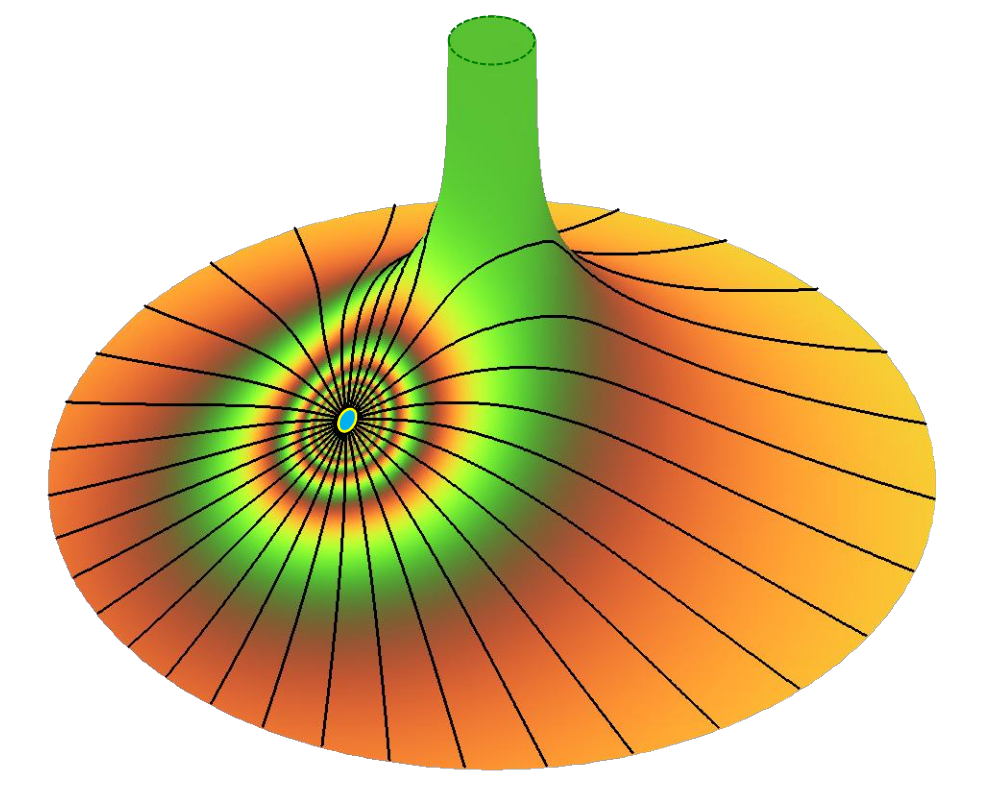}
\caption{\label{figure}This figure depicts the scalar field force
lines created by a charge in the extremal Reissner-Nordstr\"om black
hole. The surface represents the embedding of the geometry of the
equatorial section of the black hole to a three-dimensional flat
space. Color function corresponds to the value of the
scalar field $\varphi$, so that the lines of the same color
would correspond to the equipotential lines. Near the charge the
potential monotonically grows to infinity.}
\end{figure}

The static scalar Green function then reads
\be\label{GreenRN}
{\cal G}(\boldsymbol{x},\boldsymbol{x'})={\Gamma\left({n\over
2}\right)\over
4\pi^{1+{n\over
2}}}\, \cdot {1\over R^n}\,,
\ee
where the function $R$ in the isotropic spherical coordinates
$(\rho,\boldsymbol{\theta})$
takes the form
\be\begin{split}\label{R2rho}
R^2&=\rho^2+{\rho'}^2-2\rho\rho'\cos\lambda  \,, \\
\cos\lambda&=\cos\theta\cos\theta'+\sin\theta\sin\theta'\,\cos\lambda_
{n}  \,, \\
\cos\lambda_n&=\cos\theta_n\cos\theta_n'+\sin\theta_n\sin\theta_n'\,\cos\lambda_
{n-1}  \,, \\
\theta_1&=\phi\hh\theta_{n+1}=\theta\hh \lambda_{n+1}=\lambda\,.
\end{split}\ee
The meaning of the functions $\lambda$ is the length of the arc of a great
circle connecting
two points
$\boldsymbol{\theta}=(\theta_1,\dots,\theta_n,\theta)$ and
$\boldsymbol{\theta'}=(\theta'_1,\dots,\theta'_n,\theta')$ on the surface of
$n+1$-dimensional
unit sphere. The same distance expressed in terms of the curvature coordinates
$(r,\boldsymbol{\theta})$ (see \eq{RN}) reads
\be\begin{split}\label{R2r}
R^2&=\left(r^n-r_{\ins{g}}^n\right)^{2/n}
+\left({r'}^n- r_{\ins{g}}^n\right)^{2/n}  \\
&-2\left(r^n- r_{\ins{g}}^n\right)^{1/n}\left({r'}^n-
r_{\ins{g}}^n\right)^{1/n}\cos\lambda  \,, \\
\cos\lambda&=\cos\theta\cos\theta'+\sin\theta\sin\theta'\,\cos\lambda_{n}\,.
\end{split}\ee

If the charge $q'$ is
located at the point $\boldsymbol{x'}$ then the corresponding scalar field is
given by the formula \eq{varphi}
\be\label{varphi}
\varphi(\boldsymbol{x})=4\pi q'\,U^{-1}(\boldsymbol{x}')\,{\cal
G}(\boldsymbol{x},{\boldsymbol{x}'})\,.
\ee

This special case reproduces the result by Linet \cite{Linet:05} for the scalar
field near higher-dimensional extremal Reissner-Nordstr\"{o}m black hole.


\subsubsection{Scalar field in the spacetime of compactified extremally charged
black hole}

When written in the coordinates \eq{MP} the equation for the static Green
function of the scalar field does not depend on the function $U$. Therefore, on
a compact space one can independently derive the corresponding Green function
${\cal G}_\ins{(L)}$ by a similar summation over images of a scalar charge
\be
{\cal G}_{\ins{(L)}}(\boldsymbol{x},\boldsymbol{x'})=\sum_{k=-\infty}^{\infty}
{\cal
G}(\boldsymbol{x},\boldsymbol{x'}+k\boldsymbol{L})\,,
\ee

\be\label{Green_L}
{\cal G}_\ins{(L)}(\boldsymbol{x},\boldsymbol{x'})={\Gamma\left({n\over
2}\right)\over
4\pi^{1+{n\over
2}}}\, \cdot \sum_{k=-\infty}^{\infty}{1\over R_k^n}\,,
\ee
where
\be
R_k(\boldsymbol{x},\boldsymbol{x'})=|\boldsymbol{x}-\boldsymbol{x'}-k\boldsymbol
{L}|\, .
\ee
In 5D spacetime we get
\be\label{Green_L5}
{\cal G}_\ins{(L)}(\boldsymbol{x},\boldsymbol{x'})={1\over 4\pi\,{\cal
L}L}\,{\sinh{(2\pi {\cal L}/
L)}\over\cosh{(2\pi {\cal L}/ L)}-\cos{(2\pi (z-z')/ L)}}\,.
\ee
where
\be\begin{split}
z&\equiv x^{n+2}\hh
z'\equiv {x'}^{n+2}\hh
\boldsymbol{L}=(0,\dots,0,L) \,,  \\
{\cal L}^2&\equiv \delta_{ij}(x^i-{x'}^i)(x^j-{x'}^j)\hh
i,j=(1,\dots,n+1)\, .
\end{split}\ee

\section{Maxwell field of pointlike charge in Majumdar-Papapetrou spacetimes}

\subsection{Reduction of the field equations}

The Maxwell equations
\be
F^{\mu\epsilon}{}_{;\epsilon}=4\pi J^{\mu}
\hh
F_{\alpha\beta}=\nabla_{\alpha}A_{\beta}-\nabla_{\beta}A_{\alpha}\,,
\ee
for a static source $J^{\mu}=\delta^{\mu}_0 J^0$ reduce to a single equation for
the potential $A_0$
\be
{1\over\sqrt{-g}}\,\partial_{\epsilon}\left(\sqrt{-g}g^{00}g^{\epsilon\beta}
\partial_{\beta}A_0\right)=-4\pi J^0\,,
\ee
or
\be
\delta^{ab}\,\partial_{a}\left(U^{2}\partial_{b}A_0\right)=4\pi U^{2/n}\,J^0\,.
\ee
Substituting
\be
A_0=-U^{-1}\psi\,,
\ee
we obtain
\be
-\psi\lap U+U\lap \psi=-4\pi U^{2/n}\,J^0\,,
\ee
or
\be\label{psi}
\left[\lap -({U^{-1}\lap U})\right] \psi=-4\pi U^{-1+{2\over n}}\,J^0\,.
\ee
For a point-like static source with the total charge $e'=(4\pi)^{-1}$ located at
the point
$\boldsymbol{x'}$
\be
\sqrt{-g}\,J^0=U^{{2\over n}}\,J^0={1\over
4\pi}\,\delta^{n+2}(\boldsymbol{x}-\boldsymbol{x'})\,,
\ee
\be\label{Lap_psi}
U\lap\psi-\psi\lap U=-\delta^{n+2}(\boldsymbol{x}-\boldsymbol{x'})\,,
\ee
and the vector potential $A_0=-U^{-1}\psi$ gives the static Maxwell Green
function ${\cal G}_{00}(\boldsymbol{x},\boldsymbol{x'})$.
The function $\psi$ satisfies \eq{psi} which is different from the
scalar case because it contains extra $\delta$-function-like potential
terms. Though these potential terms are localized on the black hole horizons
and are multiplied by an extra  factor $U^{-1}$ which vanishes on the horizon,
one has to be careful in
dealing with this equation because  $\psi$ itself may diverge on the horizon.

\subsection{Field of a pointlike charge: general case}

Considering the class of functions $\psi$ which are not necessarily regular on
the horizon, it is suggestive to look for a solution of the form
\be\begin{split}\label{psi_0}
\psi(\boldsymbol{x},\boldsymbol{x'})&={1\over
U(\boldsymbol{x'})}{\Gamma\left({n\over 2}\right)\over 4\pi^{1+{n\over 2}}}
\left[{1\over R^n} 
+B(\boldsymbol{x},\boldsymbol{x'})\right]\,,\\
B(\boldsymbol{x},\boldsymbol{x'})&=b+\sum_i {C_i(\boldsymbol{x'})\over
\rho_i^n}\,.
\end{split}\ee
It is easy to check that
\be
\lap\psi=-{1\over
U(\boldsymbol{x'})}\left[\delta^{n+2}(\boldsymbol{x}-\boldsymbol{x'})+\sum_i
C_i(\boldsymbol{x'})
\,\delta^{n+2}(\boldsymbol{x}-\boldsymbol{x}_i)\right]
\ee
and
\be
U=1+\sum_k {M_k\over \rho_k^n}
\hh
\lap U =-{4\pi^{1+{n\over 2}}\over \Gamma\left({n\over 2}\right)}\sum_k
M_k\,\delta^{n+2}(\boldsymbol{x}-\boldsymbol{x}_k)\,.
\ee
Substitution of these relations to \eq{Lap_psi} gives the constraints on the
$C_i(\boldsymbol{x'})$
\be\begin{split}
&-\left\{1+\sum_k {M_k\over
\rho_k^n}\right\}\\
&\times\left\{\delta^{n+2}(\boldsymbol{x}-\boldsymbol{x'}
)+\sum_i
C_i(\boldsymbol{x'})
\,\delta^{n+2}(\boldsymbol{x}-\boldsymbol{x}_i)\right\}\\
&+\left\{{1\over R^n}+b+\sum_i {C_i(\boldsymbol{x'})\over
\rho_i^n}\right\}\left\{\sum_k
M_k\,\delta^{n+2}(\boldsymbol{x}-\boldsymbol{x}_k)\right\}
\\
&=-U(\boldsymbol{x'}
)\delta^{n+2}
(\boldsymbol{x}-\boldsymbol{x'})\,.
\end{split}\ee
Taking into account the identity
\be
{1\over R^n}\delta^{n+2}(\boldsymbol{x}-\boldsymbol{x}_k)={1\over
{\rho'}_k^{\,n}}\delta^{n+2}(\boldsymbol{x}-\boldsymbol{x}_k)
\ee
we obtain the condition
\be
C_k=M_k\left({1\over {\rho'}_k^n}+b\right)\,.
\ee
The meaning of this condition is that the vector potential $A_0$ has only one
pole, which is located at the position of a test charge. All other poles located
at $\boldsymbol{x}_k$, which may appear in the case of arbitrary $C_k$, have to
cancel each other. So that the charges of the black holes remain the same as
in the original background Majumdar-Papapetrou spacetime.

Thus
\be
B(\boldsymbol{x},\boldsymbol{x'})=b\, U(\boldsymbol{x})+\sum_k{M_k\over
\rho_k^n{\rho'}_k^n}\,.
\ee
Because the vector potential $A_0=-\psi/U$, we obtain the static Green
function for the Maxwell field
\be\begin{split}
{\cal G}_{00}(\boldsymbol{x},\boldsymbol{x'})&=-{\Gamma\left({n\over
2}\right)\over
4\pi^{1+{n\over 2}}}
\cdot{\left[\displaystyle{1\over R^n}+\sum_k
\displaystyle{M_k\over \rho_k^n{\rho'}_k^{\,n}}\right]\over
U(\boldsymbol{x})U(\boldsymbol{x'})}\\
&-b{\Gamma\left({n\over
2}\right)\over
4\pi^{1+{n\over 2}}}
\cdot{1\over U(\boldsymbol{x'})}\,.
\end{split}\ee
The last term does not depend on $\boldsymbol{x}$ and, hence, it is a pure
gauge. We fix it by the
requirement $A_{0}(\boldsymbol{x},\boldsymbol{x'})\rightarrow 0$ when
$x^a\rightarrow\infty$. It leads to $b=0$.
Finally we get
\be\label{G00}
{\cal G}_{00}(\boldsymbol{x},\boldsymbol{x'})=-{\Gamma\left({n\over
2}\right)\over
4\pi^{1+{n\over 2}}}
\cdot{1\over U(\boldsymbol{x})U(\boldsymbol{x'})}\left[{1\over R^n}+\sum_k
{M_k\over
\rho_k^n {\rho'_k}^n}\right]\,,
\ee
\be
U(\boldsymbol{x})=1+\sum_k {M_k\over \rho_k^n}
\hhh
\rho_k=|\boldsymbol{x}-\boldsymbol{x}_k|
\hhh
R=|\boldsymbol{x}-\boldsymbol{x}'|\,.
\ee

The total vector potential, which includes the contribution \eq{calA} of charged
black holes and of a test electric charge,  is the sum
\be
{\cal A}_0(\boldsymbol{x})+A_0(\boldsymbol{x})= \sqrt{{n+1\over
2n}}\,U^{-1}(\boldsymbol{x}) + 4\pi e'\,{\cal
G}_{00}(\boldsymbol{x},{\boldsymbol{x}'})\,.
\ee
The charge of a test particle is assumed to be much less than the charges of
the black holes. The back reaction of the spacetime on the presence of
the test
charged particle is considered to be negligible.

\subsection{Special cases}

\subsubsection{Higher dimensional extremally charged black hole}

For a single higher dimensional extremally charged black hole (see \eq{rnrho}
and \eq{RN}) one has
\be
U(\boldsymbol{x})=1+{r^n_\ins{g}\over\rho^n}=\left( 1-{r^n_\ins{g}\over r^n}
\right)^{-1}\,.
\ee

The static Green function for the vector potential is
\be\begin{split}\label{G00_RN}
{\cal G}_{00}(\boldsymbol{x},\boldsymbol{x'})&=-{\Gamma\left({n\over
2}\right)\over
4\pi^{1+{n\over 2}}}
\cdot{1\over U(\boldsymbol{x})U(\boldsymbol{x'})}\left[{1\over R^n}+
{M\over \rho^n {\rho'}^n}\right]\,,\\
&=-{\Gamma\left({n\over
2}\right)\over
4\pi^{1+{n\over 2}}}\cdot\left[
{1\over U(\boldsymbol{x})U(\boldsymbol{x'})}{1\over R^n}+
{M\over r^n {r'}^n}\right]\,,
\end{split}\ee
where the function $R$ is given by the \eq{R2rho}-\eq{R2r}.

The vector potential at the point $\boldsymbol{x}$ created by the electric
charge $e'$
located at the point $\boldsymbol{x'}$ is given by
\be
A_0(\boldsymbol{x})= 4\pi e'\,{\cal
G}_{00}(\boldsymbol{x},{\boldsymbol{x}'})\,.
\ee
This formula reproduces the result by Linet \cite{Linet:05} for the Maxwell
field created by a test electric charge near higher-dimensional extremal
Reissner-Nordstr\"{o}m black hole.
The force exerting by the electric charge $e'$ on the charge $e$ is given by
\be
f_\mu=e F_{\mu 0} =4\pi ee' \,\partial_\mu
{\cal G}_{00}(\boldsymbol{x},{\boldsymbol{x}'})\,.
\ee
where the gradient is taken at the point $\boldsymbol{x}$.


\subsubsection{Maxwell field in compactified spacetimes}

In contrast to the scalar case the Green function of the Maxwell field \eq{G00}
depends on the metric function $U(\boldsymbol{x})$. However, by
construction
this
function itself is periodic
$U(\boldsymbol{x}+k\boldsymbol{L})=U(\boldsymbol{x})$ with the
same period $L$. Therefore method of images still works and
\be
{\cal
G}_{\ins{(L)}00}(\boldsymbol{x},\boldsymbol{x'})=\sum_{k=-\infty}^{
\infty}
{\cal
G}_{00}(\boldsymbol{x},\boldsymbol{x'}+k{\bf L})\,.
\ee
\be\begin{split}\label{GL00}
&{\cal
G}_{\ins{(L)}00}(\boldsymbol{x},\boldsymbol{x'})=-{\Gamma\left({
n\over
2}\right)\over
4\pi^{1+{n\over 2}}}
\cdot{1\over U(\boldsymbol{x})U(\boldsymbol{x'})}\\
&\times
\left[\sum_{k=-\infty}^{\infty}{1\over R_k^n} + M
\left(\sum_{k=-\infty}^{\infty}{ 1\over \rho_k^n}\right)\left(
\sum_{l=-\infty}^{\infty}{ 1\over {\rho'_l}^n} \right)\right]\,,
\end{split}\ee
\be\begin{split}
U(\boldsymbol{x})&=1+M\sum_{k=-\infty}^{\infty}{1\over \rho_k^n}    
\\
\rho_k&=|\boldsymbol{x}-\boldsymbol{x}_\ins{BH}-k\boldsymbol{L}|   
\hh
R=|\boldsymbol{x}-\boldsymbol{x'}-k\boldsymbol{L}|\,.
\end{split}\ee
In five dimensions $(n=2)$ we can perform summations and get explicit
formulas
\be\begin{split}
\sum_{k=-\infty}^{\infty}{1\over R_k^2}&={\pi\over {\cal L}L}\,{\sinh{(2\pi
{\cal L}/
L)}\over\cosh{(2\pi {\cal L}/ L)}-\cos{(2\pi (z-z')/ L)}}\,,\\
\sum_{k=-\infty}^{\infty}{1\over \rho_k^2}&={\pi\over \ell L}\,{\sinh{(2\pi
\ell/
L)}\over\cosh{(2\pi \ell/ L)}-\cos{(2\pi z / L)}}\,,\\
\sum_{l=-\infty}^{\infty}{1\over {\rho'}_l^2}&={\pi\over \ell' L}\,{\sinh{(2\pi
\ell'/
L)}\over\cosh{(2\pi \ell'/ L)}-\cos{(2\pi z' / L)}}\,.
\end{split}\ee
Here
\be\begin{split}
z&\equiv x^{n+2}\hh
z'\equiv {x'}^{n+2}\hh
L^a=(0,\dots,L)\,,\\
\ell^2&\equiv\delta_{ij}x^ix^j \hh {\ell'}^2\equiv\delta_{ij}{x'}^i{x'}^j\\
{\cal L}^2&\equiv \delta_{ij}(x^i-{x'}^i)(x^j-{x'}^j)\hh
i,j=(1,\dots,n+1)\, .
\end{split}\ee

The obtained exact solutions for scalar and electric fields of a static
point-like sources allows one to calculate the self-energy of particles in the
presence of one or several extremally charged black holes and an additional
force acting on charges in the presence of black holes. We are going to study
this problem in a separate paper.


\acknowledgments{
This work was partly supported  by  the Natural Sciences and Engineering
Research Council of Canada. The authors are also grateful to the Killam Trust
for its financial support. One of the authors (V.F.) thanks the Yukawa Institute for Theoretical Physics, where this work was started, for its hospitality.
}


\eject


\begin{thebibliography}{99}
\bibitem{MEMPAR} K. S. Thorne, R. H. Price, and D. A. Macdonald, Black Holes:
The Membrane Paradigm,\\ Yale Univ. Press, New Haven and London (1986)
\bibitem{fermi} E. Fermi, On the Electrostatics of a Homogeneous Gravitational
Field and on the Weight of Electromagnetic Masses,
Nuovo Cimento, {\bf 22} (1921) 176--188
\bibitem{SmithWill:1980}  A.G. Smith  and  C.M. Will,
Force On A Static Charge Outside A Schwarzschild Black Hole,
Phys. Rev. {\bf D22} (1980) 1276--1284
\bibitem{FrolovZelnikov:1982} A.I. Zelnikov, V.P. Frolov, 
Influence of gravitation on the self-energy of charged particles,\\
Sov. Phys. JETP {\bf 55} (1982) 191--198
\bibitem{Lohiya:1982} Daksh Lohiya,
Classical Selfforce On An Electron Near A Charged, Rotating Black
Hole,\\
J. Phys. {\bf A15} (1982) 1815--1823
\bibitem{ChoTsokarosWisseman:2007}D.H.J. Cho, A.A. Tsokaros, A.G. Wiseman,
The self-force on a non-minimally coupled static scalar charge outside a
Schwarzschild black hole,
Class. Quant. Grav. {\bf 24} (2007) 1035--1048
\bibitem{Copson:1928} E. T.  Copson, Electrostatics in a gravitational field,\\
Proc. R. Soc. London {\bf A 118} (1928) 184-194
\bibitem{LL} B. Leaute and B. Linet, 
Electrostatics in a Reissner-Nordstr\"{o}m spacetime,\\
Phys. Lett. {\bf A58} (1976) 5--6
\bibitem{Linet:1976} B. Linet, 
Electrostatics and magnetostatics in the Schwarzschild metric,\\
J. Phys. A  {\bf 9} (1976) 1081-1087
\bibitem{Linet:1977} B. Linet, 
Scalar or electric charge at rest in a black hole space-time,\\
C.R. Acad. Sci. {\bf A284} (1977) 215--217
\bibitem{GibbonsWarnick:2009} G.W. Gibbons  and C. M. Warnick,\\
Universal properties of the near-horizon optical geometry,\\
Phys. Rev. {\bf D79} (2009) 064031
\bibitem{Linet:05} B. Linet, Black holes in which the electrostatic
or scalar equation is solvable in closed form,\\ 
Gen. Rel. Grav. {\bf 37} (2005) 2145--2163
\bibitem{Ortin} T. Ort\'{i}n, Gravity and Strings, Cambridge Univ.Press. (2004)
\bibitem{Maeda} K.-I. Maeda  and M. Nozawa, Black hole solutions in
string theory, \\
Prog. Theor. Phys. Suppl. {\bf 189} (2011) 310-350
\bibitem{LemosZanchin:2009} Jose P.S. Lemos and  Vilson T. Zanchin,\\
Electrically charged fluids with pressure in Newtonian gravitation and general
relativity in d spacetime dimensions: Theorems and results for Weyl type
system,\\
Phys. Rev. {\bf D80} (2009) 024010
\bibitem{ChiuHoffmann:1964} Hong-Yee Chiu  and   William F. Hoffmann,
Gravitation and relativity,\\ 
W. A. Benjamin Inc., New York -- Amsterdam (1964)
\bibitem{Myers:1986} R.C. Myers, 
Higher-dimensional black holes in compactified space-times,\\
Phys. Rev. {\bf D35} (1987) 455--466
\end{thebibliography}
\end{document}